\documentclass[preprint2]{aastex}

\newcommand{\ea}{{\it et al.}}

\newcommand{\ie}{{\it i.e.}}

\newcommand{\mdot}{\dot{\mathrm{M}}}
\newcommand{\msol}{\mathrm{M}_\odot}

\newcommand{\beq}{\begin{equation}}
\newcommand{\eeq}{\end{equation}}
\newcommand{\bdm}{\begin{displaymath}}
\newcommand{\edm}{\end{displaymath}}

\slugcomment{Submitted to the Astrophysical Journal}

\begin{document}

\title{Magnetic Collimation in PNe}

\author{T. A. Gardiner, A. Frank}
\affil{Dept. of Physics and Astronomy,\\
       University of Rochester, Rochester, NY 14627-0171}

\begin{abstract}

Recent studies have focused on the the role of initially weak toroidal
magnetic fields embedded in a stellar wind as the agent for
collimation in planetary nebulae.  In these models the wind is assumed
to be permeated by a helical magnetic field in which the poloidal
component falls off faster than the toroidal component.  The
collimation only occurs after the wind is shocked at large distances
from the stellar source.  In this paper we re-examine assumptions
built into this ``Magnetized Wind Blown Bubble'' (MWBB) model.  We
show that a self-consistent study of the model leads to a large
parameter regime where the wind is self-collimated before the shock
wave is encountered. We also explore the relation between winds in the
MWBB model and those which are produced via magneto-centrifugal
processes.  We conclude that a more detailed examination of the role
of self-collimation is needed in the context of PNe studies.

\end{abstract}

\keywords{ISM: jets and outflows --- magnetic fields --- 
magnetohydrodynamics: MHD}

\section{INTRODUCTION}

For over 30 years now, a purely hydrodynamic theory of planetary
nebulae (PNe) shaping, known as the ``interacting stellar winds''
(ISW) model has been studied in considerable detail.  The basic idea
behind the ISW model is a fast, tenuous wind which overtakes a slow,
dense wind from an earlier red giant or supergiant evolutionary phase.
Applications of this interacting stellar winds model to elliptical
or bipolar PNe relies upon a generalization (GISW) which postulates
the existence of an equatorial density enhancement in the slow wind
\citep{KahnWest85,Balick87,Icke88}.  This density enhancement could
be attributed to a binary system at the origin of the out-flowing
winds \citep{Soker,Livio}.  This is especially true if the binary
system suffers common envelope evolution.  For isolated star systems
other possible explanations for the equatorial density enhancement
exist.  One physical phenomenon possible of producing slow winds
with aspherical densities is described by the ``Wind Compressed
Disk'' model of \citet{BjorkCass93} in which streamlines from a
sufficiently rapidly rotating star converge on the equatorial plane.
Another possible mechanism for producing enhanced equatorial density
distributions via an initially dipolar field is described in the
recent paper by \citet{Matt00}.

\par
As successful as this GISW model has been for describing PNe, there
exist some prominent features which appear in new, high resolution
images that can not embraced with the purely hydrodynamic model.
These features include the presence of point-symmetric knots or ansae
and jets as well as complex multi-lobed outflow structures
\citep{Balick2000,Manchado2000,Sahai00}. While the GISW model can
produce narrow jets \citep{Ickeea92,FrankMellema96,MellemaFrank97} a
large scale gaseous torus may be required.  Thus the new features seen
in observations make alternative models appear more attractive,
especially for those nebulae with collimated point symmetric
structures.  One of the most attractive alternative models which has
been proposed is based upon another generalization of the interacting
stellar winds scenario as originally described by
\citet{ChevalierLuo}.  In this process, which we will refer to as the
``Magnetized Wind Blown Bubble'' (MWBB) model, the fast wind carries a
weak magnetic field which is strengthened after passing through the
wind shock.  The nebula is thereby shaped by the latitudinal variation
in the total pressure in the hot bubble.  This model has been studied
numerically by \citet{RozFranco} and has been subsequently generalized
\citep{GarciaSegura97,GarciaSegura99} by including a latitudinal
variation in the density and velocity of the slow and fast wind based
upon the ``Wind Compressed Disk'' model of \citet{BjorkCass93}.

\par 
In this paper we critically examine assumptions built into the MWBB
model.  We have studied the model both analytically and numerically
and we show that there is a restricted regime for which the
assumptions built into the MWBB model can hold.  This regime is what
must be called the very weak field limit and we show that to date
numerical simulations which show collimation are outside of this
limit.  In $\S2$ we give a brief description of the MWBB model for the
sake of completeness and clarity.  In $\S3$ we describe some of the
previous numerical studies of the MWBB model and the results.  In
$\S4$ we study the fast wind region appropriate for the MWBB model and
show that outside of the very weak field regime the fast wind will be
magnetically collimated.  The conclusions of this section are
applicable to the MWBB model as well as its generalization
\citep{GarciaSegura97,GarciaSegura99}. In $\S5$ we discuss the 
relationship between MHD wind acceleration, collimation as well as its
relevance to the jets observed in proto-planetary nebulae (PPNe).  We
summarize the results of this paper in $\S6$.

\section{THE MWBB MODEL}

The purpose of this section is to provide an overview of key
ingredients in the MWBB model as originally presented by
\citet{ChevalierLuo}.  The model is based upon the interacting winds
picture for planetary nebulae (for a review see \citet{Frank99}).  It
is assumed that the flow can be described by four regions separated by
three surfaces: the wind shock, the contact discontinuity, and the
ambient shock in order of increasing radius.

\par 
Similar to the \citet{Parker} solar wind solution, the MWBB model
posits that the unshocked, fast wind can be described as a spherically
expanding flow with foot-points tied to a rotating stellar surface.
Applying flux conservation to this quasi-hydrodynamic model results in
a poloidal magnetic field $B_p$ which scales as $1/r^2$ and a toroidal
magnetic field $B_\phi$ which for $r\gg R_*$ scales as $1/r$, where
$R_*$ is the stellar radius.  For a radial wind with constant velocity
$v_w$ the density scales as $1/r^2$; hence, the ratio $\sigma=
B^2_\phi/(4\pi\rho v^2_w)$ is constant in the freely expanding wind.
This ratio, $\sigma$, is the main parameter for the flow.

\par 
Next consider the region between the wind shock and the contact
discontinuity, the so-called hot bubble.  It is argued that the shock
front in the stellar wind should occur at $\sim 10^{17}$ cm. Using the
scaling of the magnetic field with radius and reasonable values for
the stellar rotation rate and fast wind velocity \citet{ChevalierLuo}
argue that at radii of interest $B_\phi/B_p \gtrsim 10^2$.  Hence they
neglect the dynamical influence of the poloidal component of the
magnetic field in the hot bubble.  It is further assumed that the
total differential $\mathrm{d}\vec{v}/\mathrm{d}t$ can be neglected.
Hence, the flow in the hot bubble is steady, i.e. locally independent
of time, with streamlines given by straight lines in the poloidal
plane.  The benefit of this assumption is that the calculation of the
gas pressure and toroidal field is decoupled from the velocity and the
solution is readily obtained.

\par 
The region between the contact discontinuity and the ambient shock is
argued to be strongly radiating. Thus it can be described in the
``thin shell approximation''.  The slow wind was taken to be
spherically symmetric since \citet{ChevalierLuo} wanted solely to
study the effect of the anisotropic (magnetic) pressure in the hot
bubble in shaping the nebula.

\par 
Using a formulation by \citet{Giuliani82}, \citet{ChevalierLuo}
derived and analyzed a set of coupled time dependent equations for the
system.  By assuming self-similarity they were able to remove the time
dependence from the equations.  The resulting set of equations still
requires numerical integration, though a limiting case for small
parameters is analyzed analytically.  The result of this limiting case
suggests that deviations from spherical symmetry for the termination
shock can be expected for $\sigma \gtrsim 0.006$.  This is further
supported by the results of their numerical calculations.  To put
numbers to this, consider the situation for the PPN phase with
$\sigma=0.006$, $\mdot=10^{-7} ~\msol/yr$, and $v_w=300~km/s$ at a
radius of $r=10^{17}~cm$. This gives a toroidal magnetic field
strength of $B_\phi\approx 10~\mu G$.  (Note that in this example we
have used slower wind speeds which are appropriate to the PPN phase.)

\section{NUMERICAL STUDIES}

The first numerical simulations of the MWBB model were carried out by
\citet{RozFranco}.  Following \citet{ChevalierLuo} they neglected the 
poloidal component of the magnetic field and included only a toroidal
magnetic field in the fast wind.  Their simulations were cylindrically
symmetric, on a grid with $300~\times~500$ cells and a resolution of
about $10^{-3}$ pc in each direction.  In their study they found that
the resulting flow was fairly insensitive to the angular variation in
the toroidal magnetic field strength.  They also found that for
$\sigma<0.05$ the flows evolve in a homologous manner as one would
expect for a purely hydrodynamic ISW model. Only for $\sigma>0.05$ did
the simulations develop a strong pole-ward flow in the hot bubble.
The authors ascribe this phenomena to the magnetic tension associated
with the shock strengthened toroidal magnetic field.  The numerical
and analytic studies of the MWBB model disagree on the minimum value
of $\sigma$ necessary for magnetic shaping of the hot bubble.
\citet{RozFranco} suggest that for $\sigma<0.05$, the length scale may
simply be much larger than the simulation domain.  Other studies of
the MWBB model \citep{GarciaSegura97,GarciaSegura99} included the
presence of a gaseous torus as in the GISW model.  In these studies a
limiting value of $\sigma > .01$ was found for effective magnetic
collimation.

\par 
We note that in the MWBB model the freely expanding wind has an
unbalanced Lorentz force associated with the helical magnetic field.
This force has a component perpendicular to the wind velocity.  Hence,
on physical grounds, it is reasonable to expect that on some length
scale and for some range of field strengths, or $\sigma$, the fast
wind will collimate on its own.  The fast wind may experience
significant collimation before it interacts with the wind shock.  The
range of initial parameters over which this occurs sets the limits on
which the fast wind collimation can be reasonably neglected.  It is to
this question which we next direct our attention.

\section{FAST WIND COLLIMATION}

A basic premise of the magnetized interacting stellar wind model of
Chevalier \& Luo is that the fast wind velocity will remain
spherically symmetric and that the $B_\phi$ will scale as $1/r$ for
large $r$.  This assumption can also be seen in the numerical
simulations by noting that the radius of the ``wind-sphere'' on which
the fast wind boundary conditions are set is
$\approx10^{16}\rightarrow10^{17}$ cm.  The Lorentz force, however, is
unbalanced, with components in a direction orthogonal to the velocity
vector.  While weak fields can not appreciably change the magnitude of
the momentum, they can deflect the flow over large length scales.  The
question which remains to be answered is, ``will the fast wind
collimate on length scales of interest and how does this collimation
depend upon the parameter $\sigma$?''

\par
We approach this problem from a perturbation analysis in spherical
polar coordinates $(r,\theta,\phi)$ with the stellar rotation axis
aligned with the radial direction for $\theta=0$.  We will present two
calculations of the deflection, both of which assume purely radial
flow for the zeroth order solution.  The first calculation is
presented as a simple, concrete example and is based upon the Parker
wind as described by \citet{Brandt1970}.  This wind model treats the
wind as purely hydrodynamic and neglects the dynamical influence of
magnetic fields and rotational motion.  The second calculation
addresses concerns about the appropriateness of the analysis for small
$r$ and moderate to large field strengths and is based upon the
\citep{WeberDavis} model as described by \citep{Barnes1974}.  As we
will see, both calculations are in good agreement.

\subsection{THE PARKER WIND}

The Parker stellar wind model is characterized by the following set of
assumptions.  The gas pressure, density and wind velocity are
spherically symmetric ($v_\theta=v_\phi=0$).  The magnetic field is
given by
%
{\setlength \arraycolsep{2pt}
\begin{eqnarray*}
B_r &=& B_* \left(\frac{R_*}{r}\right)^2 \\
B_\theta &=& 0 \\
B_{\phi} &=& -B_r \left(\frac{r \Omega \sin(\theta)}{v_r}\right)
\left(1 - \frac{R_*}{r} \right)
\end{eqnarray*}
}%
where $R_*$ is the stellar radius (the base of the stellar corona),
$\Omega$ the angular velocity, and $B_*$ the poloidal component of the
magnetic field at the stellar surface.  For simplicity we take $B_*$
to be a split monopole field and assume solid body rotation.

\par
To analyze the collimation we need to solve Newton's equation for the
motion in the $\theta$-direction.  For $v_r \gg v_{\theta}$, Newton's
equation of motion in the $\theta$-direction is approximated by
\bdm
\frac{\rho v_r}{r}\frac{\partial}{\partial r}(r v_{\theta}) =
\hat{e}_{\theta} \cdot \frac{1}{c}({\bf J}\times{\bf B}) =
F^L_\theta ~.
\edm
For $\theta\ne\pi/2$ the Lorentz force in the $\theta$-direction is
given by
\bdm
F^L_\theta = \frac{-B_\phi^2 \cot\theta}{2\pi r} ~.
\edm
Since we seek a closed form solution for the wind collimation, rather
than study any particular model for the wind acceleration we note that
for $r\ge R_*$
\bdm
|B_{\phi}| \ge B_r \left(\frac{r \Omega \sin(\theta)}{V_\infty}\right)
\left(1 - \frac{R_*}{r} \right)
\edm
where $V_\infty$ is the asymptotic value of the radial velocity and we
take $B_r$ and $\Omega$ as positive.  Using this expression for the
toroidal magnetic field strength we will obtain a lower bound on the
deflection.  The equation of motion in the $\theta$-direction thus
becomes
\bdm
\frac{\partial}{\partial r}(r v_{\theta}) =
-2\sin\theta\cos\theta
\left(\frac{r^2 \Omega^2 B_r^2 }{4\pi\rho v_r V_\infty^2}\right)
\left(1 - \frac{R_*}{r} \right)^2 ~.
\edm
Next we note that the Michel velocity $V_m$
\citep{Michel69,BelcherMacGregor76} given by
\bdm
V_m^3 = \frac{r^2 \Omega^2 B_r^2 }{4\pi\rho v_r}
\edm
is a constant for steady, radial flow.  Furthermore we note that
following the definition of the parameter $\sigma$ we have
$\sigma=({V_m}/{V_\infty})^3$.
Collecting terms we can write
\bdm
\frac{\partial}{\partial r}(r v_{\theta}) = 
(-2 \sigma V_\infty \sin\theta\cos\theta)
\left(1 - \frac{R_*}{r}\right)^2 ~.
\edm
The integration of this equation is straight forward.  Applying the
boundary condition that $v_{\theta}=0$ at $r=R_*$ we obtain
\bdm
v_\theta = -2\sigma V_\infty \sin \theta \cos \theta 
\left( 1 - x^2 +2 x \ln(x) \right)
\edm
where $x=R_*/r$.  Now to study the deflection or collimation of the
fast wind we need to calculate the streamlines of the flow.  That is,
we wish to integrate the equation
\bdm
\frac{\mathrm{d}\theta}{\mathrm{d}r} = \frac{v_\theta}{r v_r}
\edm
Once again we content ourselves with a lower bound for the deflection
and substitute $V_\infty$ for $v_r$ in this expression.  This equation
can then be integrated, resulting in a relation which describes the
path taken by a particle as it moves from the point $(R_*,\theta_*)$
on the stellar surface to the point $(r,\theta)$ some distance away.
The solution is
\bdm
\frac{\tan \theta}{\tan \theta_*} = x^{2\sigma + 4\sigma x}
\exp\{ \sigma(5 -4x -x^2) \} 
\edm
where again $x=R_*/r$.  For $r \gg R_*$ this equation is well
approximated as 
\bdm
\frac{\tan \theta}{\tan \theta_*} \approx 
\textrm{e}^{5\sigma}\left(\frac{R_*}{r}\right)^{2\sigma}~.
\edm

\begin{figure}[!ht]
\plotone{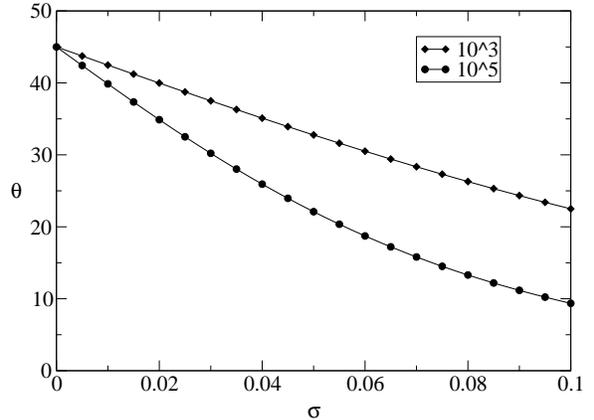}
\caption{This plot shows how the collimation of the fast wind 
depends on the parameter $\sigma$ for two radii, $(r/R_*)=10^3,~10^5$.} 
\label{angle_theta}
\end{figure}

\par
This calculation shows that the fast wind will collimate for
sufficiently high values of $\sigma$ and long propagation lengths.  In
Figure \ref{angle_theta} we plot the angle $\theta$ (in degrees) to
which the gas is deflected as a function of $\sigma$ for
$\theta_*=45^{\circ}$ and $(r/R_*)=10^3,~10^5$.  For example, for
$R_*=10^{12}$~cm Figure \ref{angle_theta} shows that by the time gas
parcels which originate at $\theta_*=45^{\circ}$ reach a radius
$r=10^{17}$ cm they will be deflected to $\theta\approx 35^{\circ}$,
$22^{\circ}$ for $\sigma=0.02$, $0.05$ respectively.

\par
We note that the Parker wind model is essentially a hydrodynamic wind
model with the magnetic field assumed weak enough to have no dynamical
influence.  Thus one might worry that the preceding argument is only
applicable for $\sigma\ll 1$.  Furthermore, it is well known that for
radii less than or equal to the Alfv\'en radius, $R_A$, the magnetic
field enforces near co-rotation of the plasma with the star.  Thus
there exits the concern that by assuming $v_\phi=0$ we have
overestimated the strength of the toroidal magnetic field at small
radii and, thereby, overestimated the degree of collimation.  In what
follows we demonstrate that collimation also occurs when the
perturbation calculation includes the plasma rotation
\citep{WeberDavis,Barnes1974}.

\subsection{THE WEBER-DAVIS WIND}

In this next calculation we will make use of the \citet{WeberDavis}
formalism.  As described by \citep{Barnes1974} we assume that
$v_\theta=0$ so that the results can be extended away from the
equator.  As in the previous calculation we will assume a split
monopole field geometry, solid body rotation and spherically symmetric
gas pressure and density for simplicity.  Furthermore, we will assume
that interior to the Alfv\'en radius that $v_\theta=0$ remains a good
approximation.  We will not assume that the rotational velocity,
$v_\phi$, is zero, but instead include it self consistently with the
toroidal magnetic field, $B_\phi$.

\par
To analyze the collimation we solve Newton's equation for the motion
in the $\theta$-direction.  For $v_r \gg v_{\theta}$, Newton's
equation of motion in the $\theta$-direction is approximated by
\bdm
\frac{\rho v_r}{r}\frac{\partial}{\partial r}(r v_{\theta}) 
-\frac{\rho v_\phi^2 \cot\theta}{r} =
\hat{e}_{\theta} \cdot \frac{1}{c}({\bf J}\times{\bf B}) =
F^L_\theta ~.
\edm
For $\theta\ne\pi/2$ the Lorentz force in the $\theta$-direction is
given by
\bdm
F^L_\theta = \frac{-B_\phi^2 \cot\theta}{2\pi r} ~.
\edm
The equation of motion in the $\theta$-direction can thus be rewritten as
\bdm
\frac{\partial}{\partial r}(r v_{\theta}) = \frac{-\cot\theta}{v_r}
\left( \frac{B_\phi^2}{2\pi\rho} - v_\phi^2 \right) ~.
\edm
Integrating the azimuthal component of the induction equation gives
\bdm
B_\phi v_r - B_r ( v_\phi -r \Omega\sin\theta) = 0
\edm
where again $\Omega$ is the angular velocity at the base of the
stellar corona.  Integrating the azimuthal component of Newton's
equation of motion gives
\bdm
v_\phi - \frac{B_r B_\phi}{4\pi\rho v_r} = \frac{L(\theta)}{r} 
\edm
where $L$ is the specific angular momentum.  At the Alfv\'en point
($r=R_A$) determined by the condition $4\pi\rho v_r^2=B_r^2$, these
two equations are linearly degenerate giving the so-called Alfv\'en
regularity condition, $L=R_A^2\Omega\sin\theta$.  Using these
relations and a bit of algebra we can solve for $v_\phi$ and $B_\phi$
in terms of parameters evaluated at the Alfv\'en radius.  The equation
of motion in the $\theta$-direction can then be rewritten in the form:
\bdm
\frac{\partial}{\partial r}(r v_{\theta}) = 
\frac{-\sin\theta\cos\theta}{v_r}
\left( \frac{2V_m^3}{v_r} (1-Y)^2 - r^2 \Omega^2 Y^2 \right) 
\edm
where $V_m$ is the Michel velocity (defined in the last section) and 
\bdm
Y = \frac{(R_A/r)^2 - \rho/\rho_A}{1 - \rho/\rho_A} ~.
\edm
The parameter $Y$ is a measure of the degree of co-rotation of the gas
due to the torque applied by the magnetic field and rotational motion
of the magnetic foot-points.  That is, $Y=0$ describes the situation
of no rotational motion, $v_\phi=0$, and $Y=1$ describes the situation
of perfect co-rotation, $B_\phi=0$.  The value of $Y$ evaluated at the
Alfv\'en radius, $Y_A$, must be understood in terms of l'H\^opital's
rule and shows that $Y_A$ is also a measure of the radial
acceleration of the wind.

\par
For $r\gg R_A$, $Y\propto (R_A/r)^2$.  Thus the equation of motion in
the $\theta$-direction shows that at large radii the force due to the
magnetic field will dominate over the centrifugal force.
Unfortunately, it is difficult to proceed further without some
knowledge of $v_r$ since this controls the radial variation of the
parameter $Y$.  One can show that for $r\ge R_A$,
\bdm
Y \le Y_A \left(\frac{R_A}{r}\right)^2 
\edm
when the radial velocity variation is slow enough that
\bdm
\frac{\mathrm{d}v_r}{\mathrm{d}r} \le
\frac{2(v_r-V_A)R_A^2}{r(r^2 - R_A^2)} ~.
\edm
It is under this condition that we can calculate a lower bound on the
collimation of the wind with radius.  The equation of motion in the
$\theta$-direction can be rewritten as:
\bdm
\frac{\partial}{\partial r}(r v_{\theta}) = 
-2\sigma V_\infty \sin\theta\cos\theta
\left( 1 - \kappa \frac{R_A^2}{r^2} \right) ~,
\edm
where
\bdm
\kappa = 2 Y_A + \frac{R_A^2 \Omega^2 Y_A^2}{2\sigma V_\infty^2} ~.
\edm
Following the same procedure described for the previous calculation
we obtain a relation for the streamline connecting the point
$(R_A,\theta_A)$ to the point $(r,\theta)$ which describes a lower
bound on the collimation.  The result is
\bdm
\frac{\tan \theta}{\tan \theta_A} = x^{2\sigma}
\exp\{ \sigma(2 + \kappa -2(1+\kappa)x + \kappa x^2) \} 
\edm
where $x=R_A/r$.  For $r \gg R_A$ this equation is well approximated
as
\bdm
\frac{\tan \theta}{\tan \theta_A} \approx 
\textrm{e}^{(2+\kappa)\sigma}\left(\frac{R_A}{r}\right)^{2\sigma} ~.
\edm

\par
For comparison with the previous calculation we would like to know,
order of magnitude, a reasonable number for the exponent
$(2+\kappa)\sigma$.  Using the definition of $\kappa$ we have
\bdm
(2+\kappa)\sigma = 2(1+Y_A)\sigma 
+ \frac{R_A^2 \Omega^2 Y_A^2}{2 V_\infty^2} ~.
\edm
To estimate the first term we need an approximation for $Y_A$.  As
described earlier, the parameter $Y$ is a measure of the co-rotation
of the gas.  Therefore, $Y_A$ is a measure of the force balance in the
$\theta$-direction at the Alfv\'en radius.  Balancing the
$\theta$-component of the Lorentz force and centrifugal force at the
Alfv\'en radius gives $Y_A=(2-\sqrt{2})$.  To estimate the second term
in the equation for $(2+\kappa)\sigma$, there are two relevant limits
which will be discussed more fully in a later section.  In the
so-called slow rotator limit the second term can be arbitrarily small.
In what is called the fast rotator limit, $V_m \approx V_\infty$ and
$V_A \approx (2/3) V_m$ \citep{BelcherMacGregor76}.  Evaluating the
Michel velocity at the Alfv\'en Radius one finds $V_m^3 = R_A^2
\Omega^2 V_A$.  Thus in the fast rotator limit we have
\bdm
(2+\kappa)\sigma = 2(1+Y_A)\sigma 
+ \frac{3 Y_A^2}{4} ~.
\edm
Inserting $Y_A=(2-\sqrt{2})$, we find the exponentials in both this
and the previous calculation to be of order $\sim 1$.  Thus we find
good agreement for the intrinsic collimation in both the Parker and
Weber-Davis model calculations.

\par
Numerical simulations which start at large radii and assume a
spherical flow are thereby constrained to small values of $\sigma$.
In order to study models with appreciable field strengths (fields
beyond the "very weak limit") requires following the flow evolution at
small radii.  This in turn necessitates including the poloidal
magnetic field since at small radii the poloidal and toroidal fields
are comparable in strength.  In summary, we conclude that a full
treatment of the propagation and magnetic collimation effects is
warranted for the study of magnetized outflows such as the MWBB model.

\section{WIND ACCELERATION AND COLLIMATION}

The collimation of magnetized winds has been extensively studied.  In
most investigations, however, both the collimation {\it and}
acceleration (also know as launching) are considered together
\citep{PudritzKonigl00}.  In view of the rather extensive literature on
the subject it is useful to attempt to put the MWBB model into the
context of these investigations.  This is particularly true because
jets often occur in the PPNe phase \citep{SahaiTrauger98,Sahai00}
before the star has become hot enough to radiatively drive a fast wind
\citep{Frank00}.  The launching of the wind is not addressed in the
MWBB model.

\par
The work of \citet{WeberDavis} was the first to explore the role of
the magnetic field in launching a wind.  This work was carried out in
1.5-D, focusing only on the launching of a wind at the equator of a
rotating star. The ability of a magnetized rotator to launch {\it
and} collimate a wind was explored in axisymmetric calculations in the
work of \citet{BlandfordPayne} and \citet{Sakurai85}.  Since that time
many workers have explored magneto-centrifugal winds (see
\citet{LamersCassinelli99} or \citet{PudritzKonigl00} for a recent 
review of the subject).  We note that a majority of the work on wind
launching and collimation has focused on the role of accretion disks
as the source of the wind.  Recently, however, their has been some
renewed interest in a star as the magnetized rotator source of the the
magnetically launched and collimated wind
\citep{BogovalovTsinganos99,TsinganosBogovalov00}.

\par
A fundamental assumption of the MWBB model is that the magnetic field
in the wind is sufficiently weak that intrinsic MHD wind collimation
can be neglected.  We now address the issue of the self-consistency of
this assumption in terms of the acceleration of the wind.
\citet{BelcherMacGregor76} utilized the Weber-Davis model to study the
transition from stellar winds in which the magnetic field is
insignificant to those in which the magnetic field is dominant in
driving the wind.  The former and later were classified as {\it slow}
and {\it fast} magnetic rotators.  This classification can also be
described by comparing the Michel velocity $V_m$ to the Parker
velocity $V_p$, where $V_p$ is defined as the velocity at infinity in
the absence of rotation.  Slow magnetic rotators have $V_m\ll V_p$,
while fast magnetic rotators have $V_m \gg V_p$.  The important point
to note here is that fast magnetic rotators drive winds via the
transfer of Poynting flux into kinetic energy flux.  In the fast
magnetic rotator limit $V_\infty \approx V_m$.

\par
Consideration of the parameters used in MWBB models shows that the
majority of the MWBB simulations carried out to date (with $\sigma <
0.1$) are not in the fast rotator regime.  (Recall that
$\sigma=({V_m}/{V_\infty})^3$.)  Thus these models are consistent with
the assumption that the wind is launched by some means other than
conversion of Poynting flux to kinetic energy flux. Conversely the
winds used in MWBB studies can not be said to have come from a form of
magnetized disk/star rotator wind model since these should have
$V_m/V_\infty \approx 1$. This is an important point as it speaks to
the origin of the wind in the PPNe stage. The results of our
perturbation analysis are of interest because they show that even a
field that can be considered weak (\ie{} $\sigma\ll1$) can produce a
pre-collimation in the wind before the wind is shocked and the MWBB
mechanism becomes operative.

\par
Finally we note that other studies of magnetic winds from slow
rotators have shown some degree of collimation.  
\citet{TsinganosBogovalov00} found that the solar wind could achieve a
significant degree of collimation at large distances when the solar
rotation rate was increased by only a factor of 5.  These results are
consistent with our perturbation analysis.

\section{CONCLUSIONS}

In this paper we've presented an analysis of the MWBB model.  We've
shown that outside of the ``very weak field'' limit, the inherent
magnetic collimation of the fast wind is neglected.  In this sense,
the numerical models which begin at large radii and include magnetic
fields beyond the very weak field limit are not consistent with the
assumed conditions at small radii.  In addition we have explored the
issue of intrinsic collimation in the MWBB winds in light of the
combined launching and collimation of winds that is possible through
magneto-centrifugal processes \citep{TsinganosBogovalov00}.  An
analysis of conditions used in MWBB studies shows that these winds can
not originate from stars lying in the fast magnetic rotator regime.
This point is relevant in determining the applicability of the MWBB
model PPNe. Many PPNe show collimated flows even though their central
stars do not have the photon momentum flux required to drive these
outflows \citep{Alc00,SahaiTrauger98}.  The role of magnetic rotators
in driving PPNe flows has recently been articulated by \citet{BFW2000}
and \citet{Blackmanea2001}.

\par
We wish to emphasize, however, that the fundamental processes
described by the MWBB model are sound and, in fact, represent a class
of behavior that has not been given adequate attention in the
literature on MHD collimation.  As the previous MWBB studies have
shown the post-shock hoop stresses in a magnetized wind with
significant toroidal field will act as an {\it additional} collimation
mechanism above and beyond the intrinsic collimation provided via the
launching process.

\par
Thus magnetohydrodynamic collimation and shaping of Planetary nebulae
as well as production of such objects as the knots and jets observed
within PNe is a very viable mechanism.  The great promise which lies
in the incorporation of magnetic fields into models of PNe formation
is in the description of the shaping mechanisms.  With this should
also come a better understanding of the period of stellar evolution
during which the mass loss is evolving from a slow to fast stellar
wind.

\end{document}